\documentclass[11pt]{article}
\usepackage{graphicx}

\begin{document}

\title{Absolutely Stable Proton and Lowering the Gauge Unification Scale}
\author{{\bf S.M. Barr} \\
{\small Department of Physics and Astronomy} \\ {\small Bartol
Research Institute,}
\\ {\small University of Delaware, Newark, Delaware 19716, USA} \\ \\
{\bf X. Calmet} \\ {\small Physics $\&$ Astronomy,} \\
{\small University of Sussex, Falmer, Brighton, BN1 9QH, UK}}
\maketitle

\maketitle

\begin{abstract}
A unified model is constructed, based on flipped $SU(5)$ in which the proton is
absolutely stable. The model requires the existence of new leptons with masses
of order the weak scale. The possibility that the unification scale could be 
extremely low is discussed.
\end{abstract}

\section{Introduction}

It is an interesting question how low the unification scale can be. 
The most obvious issue is the proton lifetime. This is not necessarily a
severe constraint on models of quark-lepton unification based on 
the Pati-Salam group \cite{PatiSalam}, since the
gauge interactions in such models do not violate baryon number. And recently the 
possibility of a very low scale for Pati-Salam unification of 
quarks and leptons has been discussed \cite{PFP-Wise}.

Proton decay is a much more serious constraint on models with gauge 
unification, by which we mean the unification of $SU(3)$ of color 
and $SU(2)$ of weak isospin in a single simple gauge group. (Gauge
unification includes both ``grand" unification and ``flipped" unification
\cite{FlippedSU5}.) A very low gauge unification scale would be conceivable,
however, if somehow the proton could be rendered absolutely stable.
Of course, the lower the unification scale the less room for the
Standard Model gauge couplings to run. Nevertheless, if there were
significant corrections to the gauge couplings 
coming from new physics at or just above the unification scale, gauge
coupling unification might nevertheless occur with a relatively low 
unification scale, perhaps even quite near the weak scale. 

In section 2 of the paper, we construct a simple model based on flipped $SU(5)$ 
\cite{FlippedSU5} in which the proton is absolutely stable due to an exact local
symmetry.   
We prove the absolute stability of the proton in the model in section 3. In section
4, we show that the symmetry that forbids proton decay in the model is 
anomaly-free and can therefore be gauged, thus ensuring that quantum 
gravity effects respect it. We also show that the model can be embedded in $SU(6)
\times SU(2)$.

In section 5, we discuss the lepton sector, which has a rich low-energy
phenomenology. The model requires
the existence of several extra lepton doublets whose masses come from the Standard 
Model Higgs field and which therefore should be accessible to accelerator
searches. Such extra lepton doublets 
would have a large effect on the amplitude for Higgs decay to two photons. 
We show how agreement with experiment is nonetheless possible. (The $H \rightarrow
\gamma \gamma$ decay rate can also be smaller or larger than the Standard Model
prediction, depending on the particle content of the
model.) In section 5, we also show
that small Majorana masses for the neutrinos can arise in a simple way despite
the absence of superheavy right-handed neutrinos in the model.

In section 6, we briefly discuss the unification of gauge couplings, and 
consider the radical possibility that the unification scale could be 
near the weak scale, and even more radically that the gravity scale
could be near the weak scale as well, eliminating all large mass hierarchies.

\section{A unified model with an absolutely stable proton}

Proton decay can be suppressed if some of the 
baryon-number-violating gauge and Higgs couplings 
to quarks and leptons
have the effect of converting $u$ or $d$ quarks 
into fermions that are too heavy to appear 
in the final state of proton decay. 
One could call this ``kinematic blocking" of 
proton decay.

The most obvious way this might happen is
that the unified interactions convert fermions of the lighter
families into those of the heavier families, for example a $d$ quark into 
a $\tau$ lepton. If the unified group is $SU(5)$ and only the Standard Model quarks
and leptons exist, then  
proton decay cannot be completely suppressed by kinematic blocking 
even at tree-level, as it was shown already in \cite{MinSU5}. On the other hand, it
was shown in
\cite{DorsnerPerez} that it can be completely 
suppressed at tree-level if the group is flipped $SU(5)$, but only by 
tuning certain fermion mixing angle within $SU(5)$ multiplets. 

If non-Standard Model quarks and leptons are introduced, more possibilities exist for kinematic blocking of proton decay.
Already in 1980 an $SU(5)$ model in which proton decay was kinematically blocked by heavy non-Standard Model fermions was proposed in \cite{SegreWeldon}. That model, however, is no longer viable, as the new quarks and leptons it proposed should have been seen by experiments. In another interesting paper \cite{bgk}, it was shown that proton decay can be forbidden to all orders in perturbation theory in $SU(5)$ by introducing into each family extra vectorlike fermions in ${\bf 5} + \overline{{\bf 5}} + 2 \times ({\bf 10} + \overline{{\bf 10}})$. However, proton stability was due to global symmetries in that model, and thus not immune to
gravitational and other non-perturbative effects. Moreover, the pattern of vacuum expectation values assumed in that model cannot be exact without fine-tuning. (Another interesting model with kinematic blocking of proton decay in a supersymmetric $SU(5)$ model  with extra dimensions is \cite{kobakhidze}.)

Recently, in \cite{SMB-rotate}, it was shown that in flipped $SU(5)$ with new non-Standard Model fermions complete kinematic blocking of proton 
decay can be achieved at tree level in a simple way without fine-tuning tuning of parameters. But even in that scheme, the
proton decay was not absolutely forbidden, since loop-diagrams and non-perturbative effects
could induce it.

The model we present here, which like that of \cite{SMB-rotate} is based on 
flipped $SU(5)$, has an absolutely
stable proton. Flipped $SU(5)$ seems to 
be the smallest unification group that allows proton 
decay to be completely forbidden by kinematic blocking in a realistic model
without any fine tuning of mixing angles. In this model, as in \cite{SMB-rotate},
the new types of heavy fermions that block proton decay
are quarks and leptons that are vectorlike under the
Standard Model group. These vectorlike fermions 
will be assumed to have masses of at least hundreds 
of GeV. We will denote these new heavy fermions by 
capital letters, and the known Standard Model fermions 
by lower-case letters.

The mechanism for blocking proton decay does not depend
on the number of families, so we will suppress family
indices throughout this paper for notational simplicity; 
and the discussion will proceed just as though there were 
only one family. It will be obvious that this does not 
matter.

The gauge group of ``flipped $SU(5)$" is
$SU(5) \times U(1)_X$, with 
the weak hypercharge $Y/2$ of the
Standard Model being a linear combination of the 
$U(1)_X$ generator $X$ and the $SU(5)$ generator
$Y_5/2 \equiv diag (\frac{1}{2}, \frac{1}{2}, - \frac{1}{3},
- \frac{1}{3}, - \frac{1}{3})$, specifically 
 $Y/2 = \frac{1}{5} (- Y_5/2 + X)$,with $X$ normalized as 
in Table I, which gives the quark and lepton content of 
the model.

\vspace{0.2cm}

\noindent {\bf Table I.} The fermion content of one family of the model. The left
column gives the $SU(5) \times U(1)_X$ quantum numbers. The other column shows how
the different species of quark and lepton are contained in the $SU(5)$ multiplets. 

\vspace{0.2cm}

\begin{tabular}{|l|lclcl|}
\hline 
$10^{(1)}$ & $\psi^{\alpha \beta}$ & =  & $ \left[ \psi^{12}, \left( 
\begin{array}{c} \psi^{1a} \\ \psi^{2a} \end{array} \right),
\psi^{ab} \right]$ & = & $\left[ \nu^c , \left( 
\begin{array}{c} u \\ d \end{array} \right),
D^c \right]$ 
\\ 
$\overline{5}^{(-3)}$ & $\psi_{\alpha}$ & = & $\left[ 
\left( \begin{array}{c} \psi_2 \\ \psi_1 \end{array} \right) , 
\psi_a \right]$
& = & $\left[ \left( 
\begin{array}{c} \nu \\ \ell^- \end{array} \right), 
u^c \right]$ 
\\
$1^{(5)}$ & $\psi$ & = & $\psi$ & = & $\ell^+$ 
\\ \hline 
$5^{(-2)}$ & $\chi^{\alpha}$ & = & $\left[ 
\left( \begin{array}{c} \chi^1 \\ \chi^2 \end{array} \right) , 
\chi^a \right]$
& = & $\left[ \left( 
\begin{array}{c} N' \\ L'^- \end{array} \right), 
D \right]$ 
\\ 
$\overline{5}^{(2)}$ & $\chi_{\alpha}$ & = & $\left[ 
\left( \begin{array}{c} \chi_1 \\ \chi_2 \end{array} \right) , 
\chi_a \right]$
& = & $\left[ \left( 
\begin{array}{c} \overline{N}'' \\ L''^+
\end{array} \right), 
d^c \right]$  
\\ \hline 
$1^{\prime (5)}$ & $\sigma'$ & & & = & $L'^+$ 
\\ 
$1^{\prime (0)}$ & $\tau'$ & & & = & $\overline{N}'$ 
\\ & & & & & \\
$1^{\prime \prime (-5)}$ & $\sigma''$ & & & = & $L''^-$ 
\\ 
$1^{\prime \prime (0)}$ & $\tau''$ & & & = & $N''$ 
\\ \hline 
\end{tabular}

\noindent The left column in Table I gives 
the $SU(5) \times U(1)_X$
representation, where the superscript is the value of $X$.
In the rest of the table, the Greek indices run from 1 to 5 and are the fundamental 
indices of $SU(5)$. The index values 1,2 correspond to the
weak hypercharge group $SU(2)_L$, while the values 3,4,5 
correspond to the QCD color group $SU(3)_c$ and are denoted by
lower-case Latin letters, $a,b, etc$. (Note, however, that 
the superscript $c$ when appearing alone denotes 
an antiparticle, for example $d^c$ denotes the 
anti-down quark.)

One sees from Table I that a single family of fermions in this model 
consists of the usual
$10^{(1)} + \overline{5}^{(-3)} + 1^{(5)}$ family of  
flipped $SU(5)$ (denoted by the letter $\psi$), 
plus a vectorlike pair $5^{(-2)} + \overline{5}^{(2)}$
(denoted by $\chi$), and a vectorlike
set of $SU(5)$-singlet fermions, denoted by $\sigma$ and $\tau$. 
This may look like
a somewhat arbitrary assortment, but, as will be seen in section 4, they
fit exactly into a small set of representations of 
$SU(6) \times SU(2)$, namely $(15,1) + (\overline{6}, 2) 
+ (1,3)$. (In fact, the smallest anomaly free chiral
set of fermions of $SU(6) \times SU(2)$ would consist
of $(15,1) + (\overline{6}, 2)$, so from the viewpoint  
of this larger group the fermions of this
model form quite an economical set.)

Notice one peculiarity of the normal flipped $SU(5)$
family shown in the first three lines of Table I: 
it contains all the fermions of a Standard Model family with
the exception of $d^c$. Where the  $d^c$ would normally
be, one finds the new heavy field $D^c$. 
And, conversely, $d^c$ occupies the place where one would
expect to find $D^c$ in the extra vectorlike pair $5^{(-2)} +
\overline{5}^{(2)}$. This
substitution is essentially what leads to the kinematic 
blocking of proton decay. How it happens will be seen 
shortly. 

This explains the need for the extra vectorlike 
$5^{(-2)} + \overline{5}^{(2)}$ of each family. What,
however,
explains the need for the extra $SU(5)$-singlet fermions (the ones denoted by
$\sigma$ and $\tau$)?
The answer is that they are there to ``mate" with the {\it leptons} in 
$5^{(-2)} + \overline{5}^{(2)}$ to give them Dirac masses.
One might ask whether these leptons could acquire mass
more simply through a mass term $M \overline{5}^{(2)}
5^{(-2)} = M \chi^{\alpha} \chi_{\alpha}$. Indeed, they could; but such a term would
necessarily 
include the term $M d^c D$, which would cause mixing between $d^c$ 
and $D^c$. That would mean that what we call $D^c$ in Table I would, through mixing, 
actually be partly the light Standard Model field. This would cause
the kinematic blocking of proton decay to be imperfect; and proton decay would
only be suppressed by a mixing angle. This is what happens in the
scheme proposed in \cite{SMB-rotate}. This would not suppress proton 
decay sufficiently if the unification scale were very low. This is the reason why
this scheme requires the existence of extra leptons whose masses come from
electroweak breaking. These leptons have many phenomenological consequences, as will
be discussed in section 5.  On the other hand, note that the extra quarks, $D$ and
$D^c$,
acquire mass from a Higgs field that is an electroweak singlet, i.e. $\Omega^{12}$, and
do not couple to the Standard Model Higgs doublet. This difference
between the extra quarks and extra leptons is a peculiar feature of the kind
of model we are discussing.

Breaking the gauge symmetries and giving quarks 
and leptons masses can be done with just the 
three types of Higgs fields shown in Table II. 

\noindent {\bf Table II.} The Higgs field content of the model.

\begin{tabular}{|l|lclcl|}
\hline $5^{(-2)}_H$ & $H^{\alpha}$ & = & $\left[ 
\left( \begin{array}{c} H^1 \\ H^2 \end{array} \right) , 
H^a \right]$
& = & $\left[ \left( 
\begin{array}{c} H^0 \\ H^- \end{array} \right), 
H^{-1/3} \right]$ 
\\ & & & & & \\
$5^{(3)}_H$ & $\tilde{H}^{\alpha}$ & = & $\left[ 
\left( \begin{array}{c} \tilde{H}^1 \\ \tilde{H}^2 \end{array} \right) , 
\tilde{H}^a \right]$
& = & $\left[ \left( 
\begin{array}{c} \tilde{H}^+ \\ \tilde{H}^0 \end{array} \right), 
\tilde{H}^{2/3} \right]$ 
\\ & & & & & \\ 
$10^{(1)}_H$ & $\Omega^{\alpha \beta}$ & = & $\left[ \Omega^{12}, \left( 
\begin{array}{c} \Omega^{1a} \\ \Omega^{2a} \end{array} \right),
\Omega^{ab} \right]$ & = & $\left[ \Omega^0 , \left( 
\begin{array}{c} \Omega^{2/3} \\ \Omega^{-1/3} \end{array} \right),
\Omega^{-2/3} \right]$  
\\ \hline
\end{tabular}

\noindent As is usual in flipped $SU(5)$ models, the breaking of
$SU(5) \times U(1)_X$ down to the Standard Model group
is done by a Higgs field that 
transforms as $10^{(1)}$, which we denote 
$\Omega^{\alpha \beta}$. The component $\Omega^{12}$ 
has $Y_5/2 = 1$ and $X = 1$ and so has $Y/2 =0$, and is 
also obviously a singlet under $SU(3)_c \times SU(2)_L$,
so that its vacuum expectation value (VEV) leaves the
Standard Model group unbroken. This VEV is of order the
unification scale. The Higgs fields that do the electroweak breaking are 
the $SU(2)_L$ doublets in the $5^{(-2)}_H$ and $5^{(3)}_H$,
which we denote respectively by $H$ and $\tilde{H}$.

There are eight Yukawa terms that are needed to give
quarks and leptons mass, which are listed in the
first seven rows of Table III, where we write
these terms using the various alternative notations
given in Table I. 

\vspace{5cm} 

\noindent {\bf Table III} The Yukawa terms needed to give mass to the fermions. The
ninth row is a term needed in the Higgs potential to align vacuum expectation
values.

\begin{tabular}{|lclcl|}
\hline $d d^c \langle \tilde{H}^* \rangle$ & = &  $\psi^{2a} \chi_a \langle
\tilde{H}_2^* \rangle$ & $\subset$ & $10^{(1)} \overline{5}^{(2)} 
\langle (5^{(3)}_H)^* \rangle$ 
\\ & & & & \\
$(u u^c, \nu^c \nu) \langle H^* \rangle$ & = & $(\psi^{1a} \psi_a, 
\psi^{12} \psi_2) \langle
H_1^* \rangle$ & $\subset$ & $10^{(1)} \overline{5}^{(-3)} 
\langle (5^{(-2)}_H)^* \rangle$ 
\\ & & & & \\
$\ell \ell^c \langle H \rangle$ & = &  $\psi_1 \psi \langle
H^1 \rangle$ & $\subset$ & $\overline{5}^{(-3)} 1^{(5)} 
\langle 5^{(-2)}_H \rangle$
\\ \hline
$D^c D \langle \Omega \rangle$ & = & $\psi^{ab} \chi^d \langle
\Omega^{12}\rangle \epsilon_{12abd}$ & $\subset$ & $10^{(1)} 5^{(-2)} 
\langle 10^{(1)}_H \rangle$
\\ & & & & \\
$L^{\prime -} L^{\prime +} \langle \tilde{H}^* \rangle$ & = &  
$\chi^2 \psi \langle
\tilde{H}_2^* \rangle$ & $\subset$ & $5^{(-2)} 1^{\prime (5)} 
\langle (5^{(3)}_H)^* \rangle$
\\ & & & & \\
$N' \overline{N}' \langle H^* \rangle$ & = &  
$\chi^1 \tau' \langle
H_1^* \rangle$ & $\subset$ & $5^{(-2)} 1'^{(0)} 
\langle (5^{(-2)}_H)^* \rangle$ 
\\ & & & & \\
$L''^- L''^+ \langle \tilde{H} \rangle$ & 
= & $\sigma'' \chi_2  \langle
\tilde{H}^2 \rangle$ & $\subset$ & $1''^{(-5)} \overline{5}^{(2)}  
\langle 5^{(3)}_H \rangle$
\\ & & & & \\
$N'' \overline{N}'' \langle H \rangle$ & 
= & $\tau'' \chi_1  \langle
H^1 \rangle$ & $\subset$ & $1''^{(0)} \overline{5}^{(2)}  
\langle 5^{(-2)}_H \rangle$
\\ \hline 
$\Omega \tilde{H}^* H^*$ & = & $\Omega^{12} \tilde{H}_2^* H_1^*$ & 
$\subset$ & $10^{(1)}_H (5^{(3)}_H)^* (5^{(-2)}_H)^*$ 
\\ \hline 
$\nu N'' \langle \tilde{H} \rangle$ & = &  
$\psi_2 \tau'' \langle \tilde{H}^2 \rangle$ &
$\subset$ & $\overline{5}^{(-3)} 1''^{(0)} \langle
5^{(3)}_H \rangle$
\\ \hline
\end{tabular}

\noindent The ninth row of Table III gives a cubic term in
the Higgs potential that is needed to tie the various 
Higgs fields together, thereby
aligning their VEVs and avoiding accidental global
symmetries in the Higgs potential that would lead to 
goldstone bosons. 

The terms in the first nine rows of Table III are needed
in the model. There are also terms that must be forbidden
if proton decay is to be suppressed. We already mentioned
one such term, namely an explicit mass term of the form
$\overline{5}^{(2)} 5^{(-2)}$, which would mix $d^c$ and 
$D^c$. Let us suppose for a moment that the terms in the
first nine rows of Table III are the {\it only} non-trivial
terms in the Yukawa sector and Higgs potential. (We count
as ``trivial" any terms that always must be present no
matter what the symmetries of the model are,
such as the absolute square of any 
Higgs field.) With only those nine terms, the model is 
easily found to have a $U(1) \times U(1)$ accidental
symmetry, which we will call $U(1)_a \times U(1)_b$.

In Table IV we give the $a$ and $b$ charges for
all the fermion and Higgs multiplets listed in Tables I and II.
These charge assignments may look somewhat random, but
it will be seen later that they have simple
group theory interpretations if the flipped $SU(5)$ 
group is embedded in $SU(6) \times SU(2)$. Moreover, note
that the $b$ values have 
a simple pattern: fermions that are odd-rank $SU(5)$ tensors have
$b = -1$, those that are even-rank tensors have $b = +1$, and 
Higgs bosons have $b = 0$.

\noindent {\bf Table IV} The $U(1)_a$ and $U(1)_B$ charges of the fields of the model. 

\begin{tabular}{|c|ccc|cc|cccc|ccc|}
\hline 
{\rm field} & $10^{(1)}$ &$\overline{5}^{(-3)}$ & $1^{(5)}$ &
$5^{(-2)}$ & $\overline{5}^{(2)}$ & $1^{\prime (5)}$ &
$1^{\prime (0)}$ & $1^{\prime \prime (-5)}$ & $1^{\prime \prime (0)}$ &
$5^{(-2)}_H$ & $5^{(3)}_H$ & $10^{(1)}_H$ \\ 
$a$ & $0$ & $-1$ & $2$ & $0$ & $1$ & $1$ & $-1$ & $-2$ & $0$ &
$-1$ & $1$ & $0$ \\
$b$ & $1$ & $-1$ & $1$ & $-1$ & $-1$ & $1$ & $1$ & $1$ & $1$ & $0$ & $0$ & $0$ 
\\ \hline
\end{tabular}

\noindent The $U(1)_a \times U(1)_b$ symmetry allows one more Yukawa
interaction, $\overline{5}^{(-3)} 1''^{(0)} \langle
5^{(3)}_H \rangle$, which is shown in the last row of Table III.
This term, which couples $\nu$ to $N''$,
has the effect of mixing of $\nu$ and  
$\overline{N}''$. As we shall see, this is harmless 
and does not destabilize the proton.
We will now prove that the symmetry $U(1)_a$ forbids all
operators of any dimension that would give proton decay.

\section{Proof of proton stability and gauging ``baryon number"}

Any effective operator that leads to proton decay must 
involve only quark and lepton fields that are lighter 
than the proton, and thus not the new vectorlike
fields denoted by capital letters in Table I. Consequently,
it can be written in the general form

\begin{equation}
\begin{array}{l}
(u,d)^m (u^c)^n (d^c)^p (\nu,\ell^-)^q (\ell^+)^r (
\overline{N}'')^s
(\langle H^0 \rangle)^t (\langle \tilde{H}^0 \rangle)^u 
(\langle \Omega \rangle)^v \\ \\
\subset (10^{(1)})^m (\overline{5}^{(-3)})^n (\overline{5}^{(2)})^p 
(\overline{5}^{(-3)})^q (1^{(5)})^r (\overline{5}^{(2)})^s
(5^{(-2)}_H)^t (5^{(3)}_H)^u 
(10^{(1)}_H)^v, 
\end{array}
\end{equation}

\noindent where the exponents $m,n,p,q,r,s,t,u,v$ are integers.
Note that here $d$ stands for either $d$ or $s$, and $e$ stands for 
either $e$ or $\mu$, since we are not showing family indices.
Also note that we have included 
$\overline{N}''$ in this product. The reason
is that this is not a purely heavy field, since (as we noted 
previously) there
is mixing between the fields that are called $\nu$ and 
$\overline{N}''$ in Table I, due to the
last term in Table IV. 

If $U(1)_a$ is not explicitly broken, then
the value of the generator $a$ of the operator in
Eq. (1) must vanish, giving

\begin{equation}
-n +p -q +2r +s -t +u = 0.
\end{equation}

\noindent By conservation of weak hypercharge, the 
value of $Y/2$ of the operator in Eq. (1) must vanish also,
giving

\begin{equation}
\frac{1}{6} m - \frac{2}{3} n + \frac{1}{3} p - \frac{1}{2} q
+ r + \frac{1}{2} s - \frac{1}{2} t + \frac{1}{2} u = 0.
\end{equation}

\noindent Multiplying Eq. (3) by 2 and subtracting Eq. (1) gives

\begin{displaymath}
\frac{1}{3} (m - n -p) = B = 0,
\end{displaymath}
  
\noindent where $B$ is baryon number. So no baryon-number-violating operators
involving only quarks and leptons lighter than the proton exist to any order.

Note that the above arithmetic shows that the linear combination of generators

\begin{equation}
\tilde{B} \equiv 2 \left( \frac{Y}{2} \right)- a
\end{equation}

\noindent is the same as baryon number for the Standard Model fermions.  (On the heavy new fermions, however, $\tilde{B}$ has peculiar values: the heavy quarks $D$ have $\tilde{B} = \frac{2}{3}$, and the heavy leptons $N'$ and $L^{'-}$ have $\tilde{B} = -1$, with the corresponding antiparticles having the opposite
values.)
As we shall see in the next section, $U(1)_a$ is anomaly free, and must be gauged
in order to protect the stability of the proton from quantum gravity effects. 
Thus, in this model we are gauging a quantum number that coincides with baryon number on the Standard Model fields. The gauging of baryon number has been discussed in other recent papers \cite{gaugingB}.

\section{Anomaly-freedom, and possible embedding in $SU(6) \times SU(2)$} 

The proof of proton stability just given assumed that the symmetry $U(1)_a$ 
is not explicitly broken. If it is a global symmetry, however, one would expect
gravitational effects 
to break it explicitly. Therefore, to render the proton absolutely stable, 
it is necessary to gauge $U(1)_a$, which would require that $U(1)_a$ be anomaly-free.
And indeed, it turns out that it is.
With the set of fermions given in Table I, and the $U(1)_a$ charges
given in Table IV, both the anomalies of $U(1)_a$ alone and 
its mixed anomalies involving $SU(5) \times U(1)_X$
vanish. That is, the following five conditions are satisfied: $Tr (a^3) = 0$, 
$Tr (a) = 0$, $Tr (a^2 X) = 0$, $Tr (a X^2) = 0$, and $Tr (a (\lambda_5)^2) = 0$ 
(where $\lambda_5$ is an $SU(5)$ generator), as
can easily be checked.

The satisfying of all these conditions seems like an amazing coincidence, 
but actually it has a simple explanation based on the group
$SU(6) \times SU(2)$. The explanation consists in these three facts: (1) the group
$SU(6) \times SU(2)$ contains $SU(5) \times U(1)_X \times U(1)_a$; (2) the set of
$SU(6) \times SU(2)$ representations $(15,1) + (\overline{6}, 2) + (1,3)$ is
anomaly-free; and (3) when decomposed under $SU(5) \times U(1)_X \times U(1)_a$ this
set of representations contains exactly the set of fermions of one family of our
model. We will now demonstrate each of these points.

That the set of $SU(6) \times SU(2)$ representations $(15,1) + (\overline{6},2)$ is 
anomaly-free is well-known and follows simply from the possibility
of embedding in $E_6$. That $(1,3)$ is anomaly-free follows simply from the fact that 
it is a real representation. Therefore, obviously, the combined set $(15,1) +
(\overline{6}, 2) + (1,3)$ is anomaly-free under $SU(6) \times SU(2)$. Moreover,
$SU(6) \times SU(2)$ 
obviously contains the subgroup $SU(5) \times U(1)_6 \times U(1)_2$, where $U(1)_6$
is the subgroup of $SU(6)$ corresponding to the diagonal generator 
$T_6 \equiv diag(\frac{1}{2}, \frac{1}{2}, \frac{1}{2}, \frac{1}{2},
\frac{1}{2}, -\frac{5}{2})$, and $U(1)_2$ is the subgroup of $SU(2)$ 
corresponding to the diagonal generator $T_2 \equiv
diag (\frac{1}{2}, - \frac{1}{2})$. If one defines $X \equiv T_6 + 5 T_2$, then the
$(15,1) + (\overline{6}, 2) + (1,3)$ is easily seen to decompose
under $SU(5) \times U(1)_X$ into exactly the set of fermions in Table I. 
And if one identifies $a$ with $2 T_2$, one immediately finds that 
those fermions have exactly the values of $a$ given in Table IV. 

It should be mentioned that if the model is embedded in $SU(6) \times SU(2)$,
then all the Yukawa couplings shown in Table III can arise from just a few types of
terms, namely terms of the form $(15,1)(15,1)\langle (15,1)_H \rangle$, $(15,1)
(\overline{6}, 2)
\langle (\overline{6}, 2)_H \rangle$, and $(\overline{6}, 2)_H (1,3) \langle (
\overline{6}, 2)_H \rangle^*$. 

The symmetry $U(1)_b$ is not contained in $SU(6) \times SU(2)$. yet it turns out,
quite remarkably, that all its anomalies vanish over the set of fermions shown in
Table I. This includes both the anomalies of $U(1)_b$ alone and its mixed anomalies
with 
$SU(5) \times U(1)_X \times U(1)_a$. This involves altogether {\it eight}
trace conditions:  $Tr (b^3) = 0$, $Tr (b) = 0$, $Tr (b^2 X) = 0$, $Tr (b X^2) = 0$,
$Tr (b^2 a) = 0$, $Tr (b a^2) = 0$, $Tr (b a X) = 0$, and $Tr (b (\lambda_5)^2) = 0$.
Because of this it is possible to gauge the full group $SU(5) \times U(1)_X \times
U(1)_a
\times U(1)_b$. However, it is not necessary to gauge $U(1)_b$ to insure
proton stability. If $U(1)_b$ is not gauged, and therefore presumably broken
explicitly by gravity effects, then several more Yukawa terms would be allowed
besides those shown in Table III. (In particular, it would allow the coupling
$\overline{5}^{(-3)} \overline{5}^{(2)} \langle 10^{(1)}_H \rangle$.)
Those additional terms would cause Standard Model 
leptons to mix with heavy, vectorlike leptons, but it can easily be shown that
it would not cause protons to decay. 

Though it is not important for proton stability that $U(1)_b$ be anomaly free, it is 
quite interesting that it is, since it involves eight independent 
non-trivial conditions, as we saw. The question is whether there is also some
underlying 
group-theoretical explanation for these cancellations based on embedding in $SU(6)
\times SU(2)$, as there
was for the anomaly-freedom of $U(1)_a$. It happens there is a partial explanation,
as we will now see.

Consider the group $SU(6) \times SU(2) \times U(1)_T$, where $(15,1)$ has $T=1$,
$(\overline{6}, 2)$ has $T = -2$, and $(1,3)$ has $T=3$. One
can easily easily check that $U(1)_T$ is anomaly-free. There are four conditions: 

\begin{equation}
\begin{array}{l}
Tr (T^3) = 15 \cdot (1)^3 + 12 \cdot (-2)^3 + 3 \cdot (3)^3 = 15 - 96 + 81 = 0, \\
Tr (T) = 15 \cdot (1) + 12 \cdot (-2) + 3 \cdot (3) = 15 - 24 + 9 = 0, \\
Tr ((T_6)^2 T) = 1 \cdot 4 \cdot (1) + 2 \cdot 1 \cdot (-2) = 0, \\
Tr ((T_2)^2 T) = 6 \cdot 1 \cdot (-2) + 1 \cdot 4 \cdot (3) = 0. 
\end{array}
\end{equation}

\noindent Though this is a surprising coincidence, it is far less surprising than the
satisfying of {\it eight} anomaly-cancellation conditions for $U(1)_b$ at the
$SU(5) \times U(1)_X$ level. If one now defines 
$b \equiv \frac{2}{3} T_6 + \frac{1}{3} T$, one discovers that the representations 
in Table I have exactly the $b$ values given in Table IV. All of this is displayed
in Table V.

\noindent {\bf Table V} How the generators of $U(1)_X$, $U(1)_a$, and $U(1)_b$ are
related to those of $SU(6) \times SU(2) \times U(1)_T$ 

\begin{tabular}{|c|r|rrr|r|r|r|}
\hline 
& & & & & $X= \;\;\;\;\;$ & $a =$ & $b = \;\;\;\;\;$ \\ 
$[SU(6) \times SU(2)]^T$ & $SU(5)$ & $T_6$ & $T_2$ & $T$ & $T_6 + 5 T_2$ &
$2 T_2$ & $\frac{2}{3} T_6 + \frac{1}{3} T$   
\\ \hline 
$(15,1)^1$ & 10 & 1 & 0 & 1 & 1 & 0 & 1 
\\  
" & 5 & $-2$ & 0 & 1 & $-2$ & 0 & $-1$ 
\\ \hline
$(\overline{6}, 2)^{-2}$ & $\overline{5}$ & $-\frac{1}{2}$ & $\frac{1}{2}$ & $-2$ 
& 2 & 1 & $-1$ 
\\ 
" & 1 & $\frac{5}{2}$ & $\frac{1}{2}$ & $-2$ 
& 5 & 1 & 1 
\\   
" & $\overline{5}$ & $-\frac{1}{2}$ & $-\frac{1}{2}$ & $-2$ 
& $-3$ & $-1$ & $-1$ 
\\ 
" & 1 & $\frac{5}{2}$ & $-\frac{1}{2}$ & $-2$ 
& 0 & $-1$ & 1
\\ \hline 
$(1,3)^3$ & 1 &  0 & 1 & 3 & 5 & 2 & 1 
\\ 
" & 1 & 0 & 0 & 3 & 0 & 0 & 1 
\\  
" & 1 & 0 & $-1$ & 3 & $-5$ & $-2$ & 1
\\ \hline
\end{tabular}

As noted above, the group $U(1)_a$ must be local 
to prevent gravity-induced proton decay. If $U(1)_a$ is gauged, however, 
its gauge boson could create difficulties. For if $U(1)_a$ is not 
spontaneously broken there is a new long-range force, while if it is 
spontaneously broken the proof of proton stability could be invalidated,
since it depended on conservation of $a$. 

First, consider the case that $SU(5) \times U(1)_X \times U(1)_a$ is
not embedded in $SU(6) \times SU(2)$. Then one can simply introduce a 
scalar field $\eta$ which is neutral under $SU(5) \times U(1)_X$
but has charge $a({\eta}) \neq 0$ under $U(1)_a$. This field can obtain a vacuum 
expectation value that makes the mass of the $U(1)_a$ gauge boson
large enough to avoid conflict with experiment. (There is also
nothing to prevent the gauge coupling of $U(1)_a$ being small.)
The existence of such a field would modify Eqs. (1) and (2). One
must put into the operator of Eq. (1) a factor $(\langle \eta \rangle)^w$,
where $w$ is some integer. Then an additional term $w a({\eta})$ would appear on
the left-hand side of Eq. (2). This changes Eq. (4) to
$B = w a({\eta})$. As long as $a({\eta})$ is not of the form $1/w$ for some integral
value of $w$, proton decay cannot happen. 

Another possibility is that $U(1)_a$ is not spontaneously broken,
but has such a tiny gauge coupling constant that the resulting long-range force
has not been seen. This seems highly implausible, but is certainly possible.

If $SU(5) \times U(1)_X \times U(1)_a$  is embedded in
$SU(6) \times SU(2)$, then the possibilities are more limited.  
The gauge coupling of $U(1)_a$ cannot then be arbitrarily small, and
the possible values of $a({\eta})$ are restricted. Moreover, if $\langle 
\eta \rangle$ is large compared to the electroweak scale, one requires that
it break $U(1)_a$ without breaking the electroweak gauge group.
The smallest $SU(6) \times SU(2)$ multiplet that has a component that
can do this is $(6,2)$.
Then $\eta^{6,2}$ (where the 6 is the $SU(6)$ index and the 2 is the 
$SU(2)$ index) has $Y/2 =0$, $I_{2L} = 0$, and $a = 1$. But this
value of $a(\eta)$ allows the proton to decay.  The smallest multiplet
that can break $U(1)_a$ above the electroweak without allowing proton decay
is a $(21,3)$ of $SU(6) \times SU(2)$. This has a component that has
$I_{2L} = Y/2 = 0$ and $a = 2$.  This allows operators that give
$\Delta B = \pm 2$, and thus possibly neutron-antineutron
oscillations, but not proton decay.

\section{The lepton phenomenology of the model}

The model presented above has six doublets of extra leptons (two for each family).
This raises several possible phenomenological problems, including consistency with
the measured value of the $\rho$ parameter, the rates for $H \rightarrow \gamma
\gamma$ and $H \rightarrow Z^0 \gamma$, and the stability of the Higgs potential. 
We shall discuss these in turn. 

The effect of the new fermions on the $\rho$ parameter can be made small if the extra 
lepton doublets are not ``split", i.e. if the neutral and charged components have
the same or nearly the same mass. This seems somewhat
artificial, but perhaps could be enforced by some symmetry, though we have not
investigated this possibility.

The extra lepton doublets will definitely contribute very significantly
to the amplitude for $H \rightarrow \gamma \gamma$. This process comes, as is well
known, from one-loop triangle graphs, where in the Standard Model the amplitude is
dominated 
by the $W$ boson loop and $t$ quark loop \cite{H-two-gamma,HHG}. In the model presented here, one must
include the diagrams with the extra charged leptons running around the loop.
The matrix element squared for $H \rightarrow \gamma \gamma$
is given by $|M|^2 = \frac{g^2 m_H^4}{32 \pi^2 m_W^2}
\left| \sum_i \alpha N_c e_i^2 F_i \right|^2$, where $i$ stands for the type of
particle in the loop, and $F_i$ is given (for $i$ being 
a gauge boson, fermion, or scalar, respectively) by 
$F_{gauge} = 2 + 3 \tau + 3 \tau (2 - \tau) f(\tau)$, $F_{fermion}
= - 2 \tau (1 + [1- \tau]f(\tau))$, $F_{scalar} = \tau(1 - \tau f(\tau))$, for $\tau
>1$,
where $\tau \equiv (2m_i/m_H)^2$, and $f(\tau) = (\sin^{-1} \sqrt{1/\tau})^2$.  
For the Standard Model contributions, one has $F_{SM} \cong F_W + F_t \cong
+8.4 - 1.8 = 6.6$. Since the six new charged leptons in our model must be heavy
enough not to have been seen, $\tau$ for them is large and $F_{L^{\pm}} \cong
6(-4/3) \cong -8$.  
This is larger than the Standard Model contribution and of opposite sign. If there are
additional fermions that contribute to $H \rightarrow \gamma \gamma$, the
total amplitude can be close to $-1$ times the Standard Model value, giving the same
rate.
Or, depending on the number and type of additional fermions, the rate could be
somewhat smaller or larger than the Standard Model prediction.

The process $H \rightarrow Z^0 \gamma$ \cite{H-Z-gamma,HHG} is not a difficulty for the model. The
present limits on this decay are very loose, and the contribution of charged leptons
to
it are highly suppressed, since the $Z$ coupling to the charged leptons is
proportional to the well-known factor $I_{3L} - 2 Q \sin^2 \theta_W = -\frac{1}{2} +
2 (0.23) \cong -0.04$.

The presence of six lepton doublets with $O(1)$ Yukawa couplings will give large
radiative contributions to the Higgs quartic self-coupling. However, we envision the
unification scale being much lower than it is in typical unified models, and the
unified theory may be effective theory valid only below some cutoff. If that cutoff scale is relatively low, the Higgs quartic coupling can remain positive  
below that scale. 

We conclude that the existence of the extra leptons is compatible with present limits.
One might worry, on the other hand, that the new quarks $D = D^c$ would present
phenomenological problems, for instance by substantially affecting the $H
\rightarrow 2$ gluon amplitude. However, a curious feature of our model is that the
extra $D + D^c$ 
quarks (unlike the extra leptons) do not couple to the Standard Model Higgs doublet,
but get their mass from a Standard Model singlet Higgs field ($\Omega^{12}$).
Thus they do not contribute to the Higgs decay amplitudes, the $\rho$ parameter,
or the running of the Higgs quartic coupling. Moreover, their mass could be much higher
than the weak scale.

Returning to the leptons, there remains the question of neutrino mass.
Realistic neutrino masses seem at first sight to be a problem for the model. The
first question is how those masses can be fractions of
an eV, since this would not emerge from the usual see-saw mechanisms if the
unification scale is very low. The second question is how to avoid neutrino masses
that are of
the same order as the quark and charged lepton masses.

If one looks at the Yukawa couplings allowed by 
$SU(5) \times U(1)_X \times U(1)_a \times U(1)_b$, all of which are shown in Table III,
one finds several mass terms for neutral fermions. Specifically, the second, eighth,
and
tenth lines of Table III have operators that give, respectively, operators of the
form $\nu^c \nu \langle H^* \rangle$, $N'' \overline{N}'' \langle H \rangle$, and
$\nu N'' \langle \tilde{H} \rangle$. Ignoring family indices, this gives a mass
matrix of the following form:

\begin{equation}
\left( \nu, \nu^c, N'', \overline{N}'' \right) 
\left( \begin{array}{cccc} 0 & \langle H^* \rangle & \langle \tilde{H} \rangle & 0 \\
\langle H^* \rangle & 0 & 0 & 0 \\
\langle \tilde{H} \rangle & 0 & 0 & \langle H \rangle \\
0 & 0 & \langle H \rangle & 0 \end{array} \right) 
\left( \begin{array}{c} \nu \\ \nu^c \\ N'' \\ \overline{N}'' \end{array}
\right).
\end{equation}  
 
\noindent This matrix has non-zero determinant, and the VEVs that appear in it are
of order the electroweak scale. Thus, one would not expect neutrino masses of order
a fraction of an eV unless some Yukawa couplings were extremely small. For example,
if the Yukawa coupling in the terms $\nu^c \nu \langle H^* \rangle$, and $N''
\overline{N}'' \langle H \rangle$ were of order $\epsilon$ and those in $\nu N''
\langle \tilde{H} \rangle$ were of order 1, then there would be (for each family)
one pseudo-Dirac neutrino with mass of order
the electroweak scale (composed approximately of $\nu$ and $N''$) and one
pseudo-Dirac neutrino with mass of order $\epsilon^2$ times the electroweak scale
(composed approximately of $\nu^c$ and $\overline{N}''$), as can be seen from the
form of Eq. (5). One would therefore need to have $\epsilon$ be of order $10^{-6}$
to $10^{-7}$ even for the third family. This seems contrived.

A more attractive possibility arises if there is an additional type of neutral
fermion introduced for each family. Let is call it $S$ and say that it is neutral
under $SU(5) \times U(1)_X \times U(1)_a$, but has $b = -1$. Then one can have a
coupling 
of the type $\psi^{12} S \langle (\Omega^{12})^* \rangle = \nu^c S \langle
(\Omega^{12})^* \rangle$, which is contained in  $10^{(1)} 1^{(0)} \langle
(10^{(1)}_H)^* \rangle$. This term is invariant under
$SU(5) \times U(1)_X \times U(1)_a \times U(1)_b$. The mass matrix then has the form

\begin{equation}
\left(S,  \nu, \nu^c, N'', \overline{N}'' \right) 
\left( \begin{array}{ccccc} 0 & 0 & \langle \Omega \rangle  & 0 & 0 \\
0 & 0 & \langle H^* \rangle & \langle \tilde{H} \rangle & 0 \\
\langle \Omega \rangle & \langle H^* \rangle & 0 & 0 & 0 \\
0 & \langle \tilde{H} \rangle & 0 & 0 & \langle H \rangle \\
0 & 0 & 0 & \langle H \rangle & 0 \end{array} \right) 
\left( \begin{array}{c} S \\ \nu \\ \nu^c \\ N'' \\ \overline{N}'' \end{array}
\right).
\end{equation} 

\noindent This matrix has one zero eigenvalue. So (for each family) there is a
massless neutral fermion.
These can be given tiny masses in various ways, one of which we will describe shortly.
These light neutrinos are linear combinations of $S$, $\nu$ and $\overline{N}''$, as
can be seen from Eq. (6).
Of course, to be consistent with bounds on lepton universality, these linear
combinations would have to be mostly
$\nu$; but this only requires certain ratios of Yukawa couplings in Eq. (6) to be 
of order $10^{-1}$. 

One can give a small mass to the neutrinos by a higher-dimension operator
of the form $S \; S \langle \zeta \rangle^n$, where $\zeta$ is a $1^{(0)}$ of 
$SU(5) \times U(1)_X$ and has $a = 0$ and $b = 1/n$. Such an operator might be
induced by gravity. 
The value of $n$ needed to get realistic neutrino masses would depend on the gravity 
scale.
 
\section{Low scale grand unification?}

If a unified model has an absolutely stable proton due to an exact symmetry, 
then obviously proton lifetime limits would not constrain the unification scale at all.
The question would then arise how low the unification scale could be. Could it
be near the electroweak scale? If the unification scale is low, one has to explain how
the gauge couplings are able to unify. One possibility is to exploit an idea first proposed by Shafi and Wetterich many years ago \cite{Shafi:1983gz} (see also \cite{Hall:1992kq,Calmet:2008df}).

Let us consider an effective operator for physics below some cutoff scale $\Lambda$
given by
\begin{eqnarray}
\label{dim5} 
\frac{c}{\Lambda} {\rm Tr}\left(G_{\mu\nu} G^{\mu\nu} {\cal A} \right)~,
\end{eqnarray}
where $G_{\mu\nu}$ is the Grand Unified Theory field strength and ${\cal A}$ is a
scalar multiplet in the adjoint representation of $SU(5)$. The scale $\Lambda$ is kept
as a free parameter for the time being. 
Upon symmetry breaking at the unification scale $M_U$, the Higgs field gets a vacuum
expectation value $\left\langle {\cal A} \right\rangle = M_U
\left(2,2,2,-3,-3\right)/\sqrt{50\pi\alpha_G}$, where $\alpha_G$ is the value of the
$SU(5)$ gauge coupling at $M_U$. 

The dimension 5 operator modifies the gauge kinetic terms of
$SU(3) \times SU(2) \times U(1)$ below the scale $M_u$ to
\begin{equation}
\label{gaugekineticterm}
-\frac{1}{4} \left(1+\epsilon_1\right)F_{\mu\nu} F^{\mu\nu}_{{\rm U}(1)}
-\frac{1}{2}\left(1+\epsilon_2\right){\rm Tr}\left(F_{\mu\nu} F^{\mu\nu}_{{\rm
SU}(2)}\right)\\
 -\frac{1}{2}\left(1+\epsilon_3\right){\rm Tr}\left(F_{\mu\nu} F^{\mu\nu}_{{\rm
SU}(3)}\right)
\end{equation}
with
\begin{equation}
\label{epsilons}
\epsilon_1=\frac{\epsilon_2}{3}=-\frac{\epsilon_3}{2}=\frac{\sqrt{2}}{5\sqrt{\pi}}\frac{c}{\sqrt{\alpha_G}}\frac{M_u}{\Lambda}~.
\end{equation}

\noindent If we were to take $\Lambda=M_u$, then
\begin{equation}
\epsilon_1=\frac{\epsilon_2}{3}=-\frac{\epsilon_3}{2}=\frac{\sqrt{2}}{5\sqrt{\pi}}\frac{c}{\sqrt{\alpha_G}}.
\end{equation}

We can now perform a finite field redefinition $A_{\mu}^{i} \to
\left(1+\epsilon_i\right)^{1/2} A_{\mu}^{i}$ to canonically normalize the kinetic
terms of the gauge bosons. Then the corresponding redefined coupling constants  are
$g_i \to \left(1+\epsilon_i\right)^{-1/2} g_i$. We get the unification condition:
\begin{equation}
\label{boundarycondition}
\alpha_G  = \left(1+\epsilon_1\right) \alpha_1(M_u)=\left(1+\epsilon_2\right)
\alpha_2(M_u) \\
 = \left(1+\epsilon_3\right) \alpha_3(M_u)~.
\end{equation}

We now wish to consider low scale unification. Direct observational bounds on the
heavy gauge bosons of
$SU(5)/G_{SM}$ as well as on the color octet scalars lead us to consider a
unification scale in the few TeV region.
 We thus take $M_U~$ a few TeV  which implies that there is very little running for
the gauge couplings and we can use the LEP values, at least to first approximation.
We take $\alpha_2(M_Z)=0.03322$ and  $\alpha_3(M_Z)=0.118$. Since $\alpha_1$ is a
free parameter, we will use $c$ to obtain the numerical unification of $\alpha_2$
and $\alpha_3$. We need 
\begin{equation}
c= 5\sqrt{\frac{\pi}{2}}  \left (\frac{\alpha_2-\alpha_3}{3 \alpha_2 + 2 \alpha_3 }
\right) \sqrt{\alpha_G}.
\end{equation}
We then find
\begin{equation}
\epsilon_1=5 \left (\frac{\alpha_2-\alpha_3}{3 \alpha_2 + 2 \alpha_3 } \right)
\frac{1}{\alpha_G}
\end{equation}
and thus
\begin{equation}
\alpha_1=\frac{\alpha_G}{1+5 \left (\frac{\alpha_2-\alpha_3}{3 \alpha_2 + 2 \alpha_3
} \right) \frac{1}{\alpha_G} }
\end{equation}

  Numerically we have $c=-1.58/\sqrt{\alpha_G}$. If I take $\alpha_G=0.05$ for
illustration, We get $c=-7$. In a sense we see that if the grand unified theory is
strongly coupled, the Wilsonian expansion works best as the Wilson coefficients
get smaller: for $\alpha_G=1$, we get $c=-1.58$.

The $U(1)$ of hypercharge is not purely a subgroup of $SU(5)$, but lies partly in
the $U(1)_X$, whose gauge coupling is an independent, free parameter. This coupling
can be chosen to give the observed value of the hypercharge (and electromagnetic)
gauge coupling. 

Interestingly, the Planck scale could also be lowered to the TeV region to remove
all hierarchies. There are two known mechanisms for that. 
One is to assume that large extra dimensions open up at an energy scale of a few TeV \cite{ArkaniHamed:1998rs,Randall:1999ee}. The other one relies on a large hidden sector of particles which lead to a running of the Planck mass \cite{Calmet:2008tn}. Note that the running can also be obtained by a scalar field with a large non-minimal coupling to the Ricci scalar \cite{Atkins:2010re}.

\section*{Acknowledgements}

We thank K.S. Babu, Ilia Gogoladze, and Qaisar Shafi for useful conversations.
This work is supported in part by DOE grant No. DE-FG02-12ER41808.


\begin{thebibliography}{999}
\bibitem{PatiSalam} J.C. Pati and A. Salam, {\it Phys. Rev.} {\bf D10},
275 (1974). 
\bibitem{PFP-Wise} P. Fileviez Perez and M.B. Wise, {\it Phys. Rev.} {\bf D88},
057703 (2013).
\bibitem{FlippedSU5} A. DeRujula, H. Georgi and S.L. Glashow, {\it Phys. Rev. Lett.}
{\bf 45}, 413 (1980);
H. Georgi, S.L. Glashow and M. Machacek, {\it Phys. Rev.} {\bf D23}, 783 (1981);
S.M. Barr, {\it Phys. Lett.} {\bf B112}, 219 (1982). 
\bibitem{MinSU5} C. Jarlskog, {\it Phys. Lett.} {\bf B82}, 401 (1979);
R. Mohapatra, {\it Phys. Rev. Lett.} {\bf 43}, 893 (1979);
S. Nandi, A. Stern and E.C.G. Sudarshan, {\it Phys. Lett.} {\bf B113}, 165 (1982).
\bibitem{DorsnerPerez} I. Dorsner and P. Fileviez Perez, {\it Phys. Lett.} {\bf
B606}, 367 (2005).
\bibitem{SegreWeldon}  G. Segre and H.A. Weldon. {\it Phys. Rev. Lett.} {\bf 44},
1737 (1980).
\bibitem{bgk} Z. Berezhiani, I. Gogoladze and A.B. Kobakhidze,
{\it Phys. Lett.} {\bf B522}, 107 (2001). 
\bibitem{kobakhidze} A.B. Kobakhidze, {\it Phys. Lett.} {\bf B514}, 131 (2001).
\bibitem{SMB-rotate} S.M. Barr, {\it Phys. Rev.} {\bf D88}, 057702 (2013).
\bibitem{gaugingB} P. Fileviez Perez and M.B. Wise, {\it JHEP} {\bf 1108}, 068 (2011);
M. D\"{u}rr, P. Fileviez Pérez and M.B. Wise, {\it Phys. Rev. Lett.} {\bf 110}, 23 (2013)
231801.
\bibitem{H-two-gamma} J. Ellis, M.K. Gaillard and D.V. Nanopoulos, {\it Nucl. Phys.} 
{\bf B106}, 292 (1976); M.A. Shifman, A.I. Vainshtein, M.B. Voloshin and
V.I. Zakharov, {\it Sov. J. Nucl. Phys.} {\bf 30}, 711 (1979);
L. Bergstrom, G. Hulth and H. Snellman,
{\it  Z. Phys.}{\bf C16}, 263 (1983).
\bibitem{HHG} J. Gunion, H. Haber, G. Kane and S. Dawson, {\it The Higgs Hunter's
Guide} (Addison-Wesley, Reading, 1990).
\bibitem{H-Z-gamma} R.N. Cahn, M.S. Chanowitz and N. Fleishon, {\it Phys. Lett.}
{\bf B82}, 113 (1979).
\bibitem{Shafi:1983gz} 
Q. Shafi and C. Wetterich,
{\it Phys. Rev. Lett.}  {\bf 52}, 875 (1984).
\bibitem{Hall:1992kq} 
L.J. Hall and U. Sarid,
{\it Phys. Rev. Lett.} {\bf 70}, 2673 (1993).
\bibitem{Calmet:2008df} 
X. Calmet, S.D.H. Hsu and D. Reeb,
{\it Phys. Rev. Lett.} {\bf 101}, 171802 (2008).
\bibitem{ArkaniHamed:1998rs} 
N. Arkani-Hamed, S. Dimopoulos and G.R. Dvali,
{\it Phys. Lett.} {\bf B429}, 263 (1998);
I. Antoniadis, N. Arkani-Hamed, S. Dimopoulos and G.R. Dvali,
{\it Phys. Lett.} {\bf B436}, 257 (1998).
\bibitem{Randall:1999ee} 
L. Randall and R. Sundrum,
{\it Phys. Rev. Lett.}  {\bf 83}, 3370 (1999).
\bibitem{Calmet:2008tn} 
X. Calmet, S.D.H. Hsu and D. Reeb,
{\it Phys.\ Rev.\ D} {\bf 77}, 125015 (2008).
\bibitem{Atkins:2010re} 
M. Atkins and X. Calmet,
{\it Eur. Phys. J.} {\bf C70}, 381 (2010).
\end{thebibliography}
\end{document}